\documentclass[aps,pre,showpacs,floatfix,superscriptaddress,twocolumn,color]{revtex4}

\usepackage{graphicx}
\usepackage{amsmath}
\usepackage{amsfonts}
\usepackage{amssymb}
\usepackage{wasysym}
\usepackage{color}
\usepackage[normalem]{ulem}
\usepackage{hyperref}
\usepackage{xcolor}
\hypersetup{
  colorlinks   = true,	
  urlcolor     = red,	
  linkcolor    = blue,	
  citecolor   = blue	
}
\usepackage{enumitem}
\usepackage[caption=false]{subfig}

\definecolor{rmarker}{rgb}{0.9,0.0,0.0}
\definecolor{cmarker}{rgb}{0.0,0.7,0.7}

\begin{document}

\title{Mobility-induced persistent chimera states}

\author{Gabriela Petrungaro}
\affiliation{Instituto de Investigaci\'on en Biomedicina de Buenos Aires (IBioBA)\---CONICET\---Partner Institute of the Max Planck Society, Polo Cient\'{\i}fico Tecnol\'ogico, Godoy Cruz 2390, C1425FQD, Buenos Aires, Argentina}
\affiliation{Departamento de F\'{\i}sica, FCEyN UBA, Ciudad Universitaria, 1428 Buenos Aires, Argentina}%
\author{Koichiro Uriu}
\affiliation{Graduate School of Natural Science and Technology, Kanazawa University, Kakuma-machi, Kanazawa 920-1192, Japan}
\author{Luis G. Morelli}	\email{lmorelli@ibioba-mpsp-conicet.gov.ar}
\affiliation{Instituto de Investigaci\'on en Biomedicina de Buenos Aires (IBioBA)\---CONICET\---Partner Institute of the Max Planck Society, Polo Cient\'{\i}fico Tecnol\'ogico, Godoy Cruz 2390, C1425FQD, Buenos Aires, Argentina}
\affiliation{Departamento de F\'{\i}sica, FCEyN UBA, Ciudad Universitaria, 1428 Buenos Aires, Argentina}%
\affiliation{Max Planck Institute for Molecular Physiology, Department of Systemic Cell Biology, Otto-Hahn-Str.~11, D-44227 Dortmund, Germany}%
\date{\today}

\begin{abstract}
We study the dynamics of mobile, locally coupled identical oscillators in the presence of coupling delays.
We find different kinds of chimera states, in which coherent in-phase and anti-phase domains coexist with incoherent domains.
These chimera states are dynamic and can persist for long times for intermediate mobility values.
We discuss the mechanisms leading to the formation of these chimera states in different mobility regimes.
This finding could be relevant for natural and technological systems composed of mobile communicating agents.
\end{abstract}
\maketitle

\section{Introduction} \label{sec.intro}
Coupled oscillators give rise to collective dynamics in many natural and technological contexts~\cite{winfree, kuramoto, pikovsky, manrubia}.
In biological systems, coupled oscillators play a crucial role in self organising collective rhythms, 
for example in cardiac tissue~\cite{nitsan16}, circadian rhythms~\cite{herzog07, muraro13} and 
the vertebrate segmentation clock~\cite{hubaud14, harima14}.
Collective rhythms can emerge from the 
synchronization of a population of coupled individual oscillators~\cite{kuramoto, pikovsky}.
A system of coupled oscillators involves multiple timescales, determined by autonomous frequencies, 
coupling strength, shear, noise strength, coupling delay and the rate of movement of oscillators.
%
The interplay between timescales in systems of coupled oscillators may result in complex dynamics and 
has motivated the field to search for new and interesting dynamic phenomena.
For example, the interplay between coupling strength and frequency diversity is behind the paradigmatic synchronization transition~\cite{winfree, kuramoto, strogatz00}.
Shear diversity also competes with coupling strength in this transition~\cite{montbrio11}.
The relation of coupling delay and coupling strength determines a shift in collective frequency and multistability~\cite{schuster89}, 
and can affect pattern formation~\cite{ares12}.
The interplay of coupling strength and mobility sets different regimes of synchronization dynamics from local to mean field behavior~\cite{uriu10a, uriu13}.

Two key timescales of coupled systems whose interplay has not been explored are those related to coupling delay and mobility.
Coupling delays can result from the complexity of the communication mechanism, leading to finite characteristic times for either sending or processing signals.
Coupling delays are ubiquitous in cellular systems~\cite{lewis03, herrgen10} and 
can profoundly affect dynamics~\cite{schuster89,yeung99} and pattern formation~\cite{niebur91, jeong02, morelli09, ares12}.
Mobility of oscillators sets the timescale for how often single oscillators exchange neighbors, which is particularly relevant in locally coupled systems.
Mobility has been shown to reduce the time the system needs to achieve synchronization~\cite{uriu10a, uriu12, uriu13, uriu14a}, 
by extending the effective range of the coupling~\cite{uriu13} or through coarsening~\cite{levis17}.
Thus, both coupling delay and mobility independently have distinct effects on the dynamics of coupled oscillators.

An interesting question is how the different timescales of coupling delay and mobility interact 
and what is their impact on oscillatory dynamics and collective organization.
Due to coupling delay, information arriving at one oscillator at the present time was sent at a previous time from another oscillator.
The oscillator that sent the signal was close at the time of the interaction but can now be elsewhere due to mobility.
In this paper we study a model that incorporates these two timescales.
In contrast to the expectation that mobility favors the relaxation to homogeneous states~\cite{reichenbach07, uriu10a, uriu13}, 
here we find that when considered together with coupling delays mobility can also drive the system into heterogeneous states with complex long lived patterns.

\section{Theory}
%

We consider a system of $N$ identical phase oscillators placed in a one-dimensional lattice of $N$ sites.
Oscillators can move through the lattice by exchanging positions with their nearest-neighbors.
The stochastic exchange of positions is modeled as a Poisson process.
We introduce a mobility rate $\lambda$ such that each pair of neighboring oscillators exchange positions with a probability $\lambda /2$ per unit time~\cite{uriu10a, uriu13, uriu14a}.
%
With this modeling, the waiting time for the next exchange event for each oscillator is stochastic and its statistics obey an exponential distribution with mean $1/\lambda$.
%
%

The state of oscillator $i$, with $i=1, \ldots, N$, is described by a phase $\theta_i(t)$ and a position $x_i(t)$ in the lattice, 
with $x_i=1, \ldots, N$ without loss of generality. 
%
%
%
%
Position $x_i(t)$ is a discrete variable that refers to a lattice site and it is piecewise-constant. The value of $x_i(t)$ changes only when oscillator $i$ exchanges its position.
%
%
In between exchange events, phase dynamics is given by 
\begin{equation} \label{eq:model}
\dot{\theta _i}(t) = \omega 
+ \dfrac{\kappa}{n_i} \sum _{j \in V_i(t-\tau )} \sin \left( \theta _j(t-\tau) - \theta _i (t) \right)
\end{equation}
with
$$V_i (t) = \left\{j \text{ such that } |x_j(t) -x_i(t) | = 1 \right\}$$
where $\omega$ is the autonomous frequency of the oscillators, $\kappa$ is the coupling strength and $n _i$ is the number of neighbors of oscillator $i$. 
%
%
The coupling delay $\tau$ accounts for the time it takes to process a received signal.
Therefore the coupling includes a summation over the neighborhood at the time of the interaction $V_i(t-\tau)$. 
%
%
Due to mobility the neighborhood at time $t-\tau$ may be different from the one at present time $t$. 
We use open boundary conditions: the oscillators at both ends of the lattice 
interact only with their single left or right neighbors respectively and can only exchange positions with them. 
This choice of open boundary conditions prevents the formation of stable twisted states that may appear for periodic boundary conditions~\cite{wiley06, peruani10}, simplifying the analysis. 
\begin{figure}[t]
\includegraphics[width=\columnwidth]{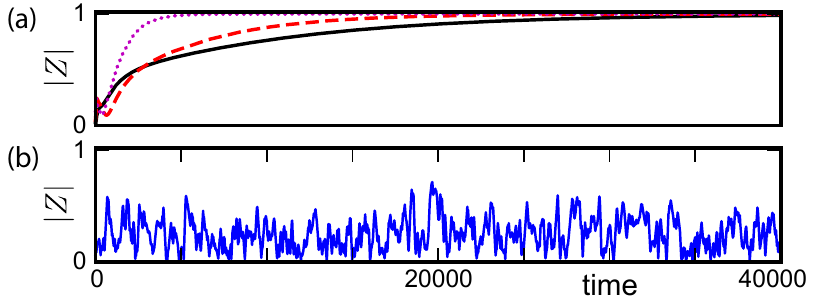}
\caption{Order parameter shows erratic behavior in the presence of mobility and coupling delay.
The modulus $|Z|$ of the order parameter as a function of time in simulations of Eq.~(\ref{eq:model}) with
(a) $\tau=0$ and $\lambda /\kappa = 0$ (solid black line), $\lambda /\kappa = 1$ (dashed red line) and $\lambda /\kappa = 10$ (dotted magenta line), and 
(b) $\tau =1.2$ and $\lambda /\kappa =1$.
In both panels $\omega =1$, $N=100$.
}
\label{fig:order_parameter}
\end{figure}

The model includes four independent timescales related to phase dynamics $\kappa^{-1}$ and $\omega^{-1}$,
mobility $\lambda^{-1}$, and coupling delay $\tau$.
The interplay of mobility and phase dynamics is characterized by the ratio $\lambda /\kappa$.
In the absence of coupling delay $\tau = 0$, the onset of the effects of mobility on collective dynamics occurs at $\lambda /\kappa=1$~\cite{uriu13}. 
In the following we set $\kappa=0.1$ and $\omega=1$ and only vary $\lambda$ and $\tau $. 
Results reported below were obtained for a lattice of $N=100$ oscillators. Initial conditions for the phase of each oscillator were chosen randomly from a uniform distribution between $0$ and $2\pi $ unless noted otherwise. We consider the interaction between oscillators starts at $t=0$ so $\theta _i(t)=\theta _i(t=0) + \omega \, t$ for $t \in [-\tau , 0]$ and $\forall i$.

We perform numerical simulations using a fourth order Runge-Kutta integration scheme.
Position exchange is implemented using an approximation of the Gillespie algorithm for discrete time intervals~\cite{gillespie76, uriu12}. For a detailed description of numerical methods see Appendix~A.

\section{Results}
In the absence of delays, mobility can speed up synchronization~\cite{uriu13}. 
Two routes to global synchronization are observed.
For low mobility, the system synchronizes by forming local order patterns that slowly relax to global synchrony.
For larger mobility a mean field behavior dominates, and synchrony is achieved without the formation of local patterns.
The time evolution of the modulus $|Z|$ of the complex order parameter
$Z(t) = N^{-1} \sum_{j=1}^{N} {e^{i \theta_j(t)}}$~\cite{pikovsky, manrubia}, 
shows that synchrony is reached much faster for larger mobility, through this second route, Fig.~\ref{fig:order_parameter}(a).
Next, we examine the time evolution of $|Z|$ in the presence of coupling delay and mobility. 
We choose a mobility rate $\lambda/\kappa=1$, which is expected to affect synchronization dynamics, Fig.~\ref{fig:order_parameter}(a)~\cite{uriu13}.
For some values of the coupling delay 
$|Z|$ exhibits a persistent erratic behavior, Fig.~\ref{fig:order_parameter}(b).
This behavior suggests that in the presence of coupling delay, mobility induces a state that differs from the known two routes.
\begin{figure}[t]
\includegraphics[width=\columnwidth]{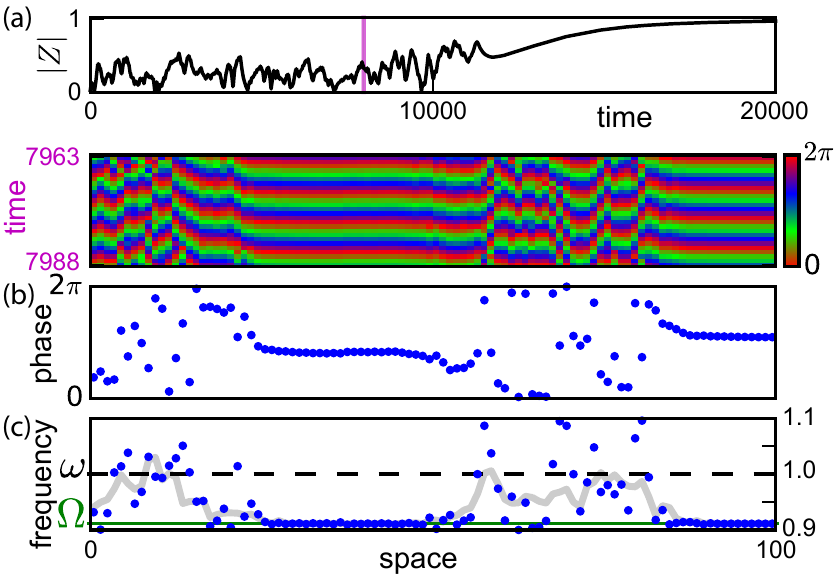}
\caption{In-phase chimera states.
(a) Modulus $|Z|$ of the order parameter vs. time for one realization in which the system exhibits chimera states (top panel).	
Space-time plot of the phases (bottom panel).
The pink vertical bar in the top panel shows the time window expanded in the bottom panel.
(b) Snapshot of a chimera state for the last frame of the time window above. 
(c) Instantaneous frequencies of oscillators (blue dots) and averaged frequencies at each lattice site (solid gray line) at the same time as in (b).
Dashed black line is the autonomous frequency $\omega $ and solid green line is collective frequency of the in-phase state $\Omega$.
Parameter values are $\lambda /\kappa =1$, $\tau =1.2$, $\omega =1$, $N=100$.
Averaged frequencies were computed with $\Delta T \, \omega / 2\pi \approx 20$.
}
\label{fig:chim_and_PO}
\end{figure}

\begin{figure}[t]
\includegraphics[width=\columnwidth]{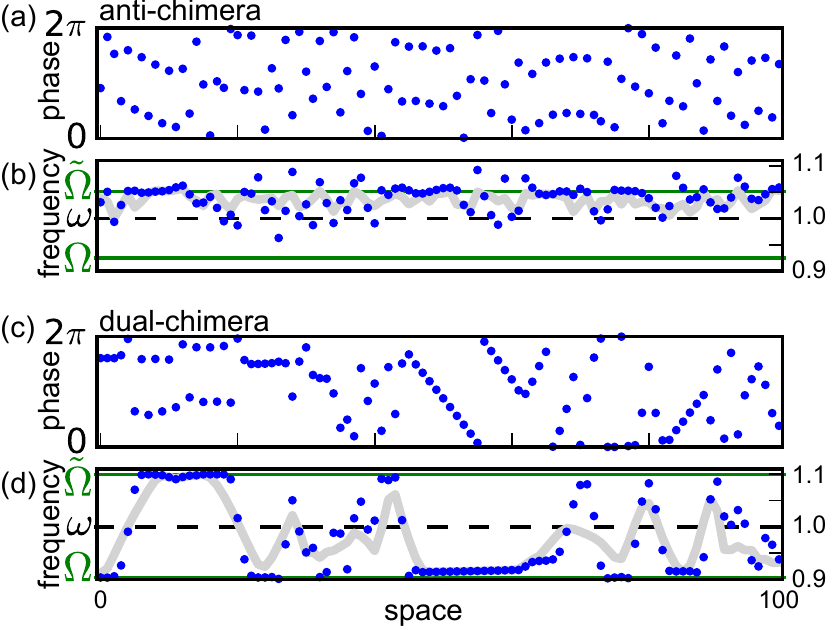}
\caption{
Anti-chimera and dual-chimera states.
Snapshots and instantaneous frequencies for (a,b) an anti-chimera state and (c,d) a dual-chimera state.
(b,d) Instantaneous frequencies of oscillators (blue dots) and averaged frequencies at each lattice site (solid gray line) at the same time as in (a,c) respectively.
Dashed black line is the autonomous frequency $\omega $ and solid green lines are the collective frequencies of the in-phase state $\Omega$ and the anti-phase state $\tilde{\Omega }$. 
Parameter values are (a,b) $\lambda /\kappa =0.1$ and $\tau =2.48$ and (c,d) $\lambda /\kappa =0.001$ and $\tau =1.44$. 
%
%
Averaged frequencies were computed with $\Delta T \, \omega / 2\pi \approx 20$.
Snapshots taken after at least $t=4000$.}
\label{fig:other_chim}
\end{figure}

Looking closer at the phase dynamics, the observed behavior of the order parameter is the consequence of complex spatio-temporal patterns, 
in which in-phase synchronous domains coexist with asynchronous domains, Fig.~\ref{fig:chim_and_PO}.
This kind of patterns were first observed in systems with non-local coupling~\cite{kuramoto02} and subsequently named \emph{chimera states}~\cite{abrams04}.
We observe that chimera states are long-lived and dynamic, Fig.~\ref{fig:chim_and_PO}(a).
Coherent domains spontaneously form out of incoherence, change their sizes and die out while other domains 
may be born in a different place in the lattice (supplementary movies S1 and S2). 
Besides in-phase chimera states where in-phase order coexists with disorder, for low mobility we also find other kinds:
anti-chimera states where anti-phase order coexists with disorder and 
dual-chimera states where both types of order coexist with disorder, Fig.~\ref{fig:other_chim}.

Instantaneous and averaged phase velocities at each lattice site show that oscillators within coherent domains have the same frequency, 
Fig.~\ref{fig:chim_and_PO}(c) and Fig.~\ref{fig:other_chim}(b,d).
The value of the frequency within coherent domains coincides with the collective frequency
of in-phase and anti-phase solutions of Eq.~(\ref{eq:model}) for non-mobile oscillators:
$\Omega = \omega - \kappa \sin(\Omega \tau)$ for in-phase~\cite{schuster89, niebur91, yeung99, jorg14, morelli09} and 
$\tilde{\Omega }= \omega + \kappa \sin(\tilde {\Omega }\tau)$ for anti-phase coherence~\cite{nakamura94, wetzel12}.
In contrast, frequencies are not locked between lattice sites in disordered domains.
Averaged frequencies over a time window $\Delta T$ allow to visualize a smoother transition between the domains, gray lines in Fig.~\ref{fig:chim_and_PO}(c) and Fig.~\ref{fig:other_chim}(b,d). 
%
%
However, chimera states are dynamic, forming, disassembling and moving across the lattice.
A larger value of $\Delta T$ will result in a smoother average frequency profile, but at the cost of blurring the boundaries between ordered and disordered domains.
%
Small coherent domains which may form and disassemble faster may not be visible in this way.

We have also observed chimera states for periodic boundary conditions (supplementary movie S3). 
Hence, although an open boundary condition breaks translation invariance at boundaries, this is not crucial for the formation of chimera states. 
Therefore, we employ open boundary conditions avoiding stable twisted states in the rest of the paper.

We seek to identify how the occurrence of chimera states in the system depends on time and parameter values.
%
%
A diversity of dynamical states, such as locally ordered states or states combining domains of in-phase and anti-phase order,
are present together with chimera states. 
%
%
To distinguish between these states we devise a method that introduces phase difference motifs to identify ordered and disordered domains in the lattice, see Appendix~B.
\begin{figure}[t]
\centering
\includegraphics[width=\columnwidth]{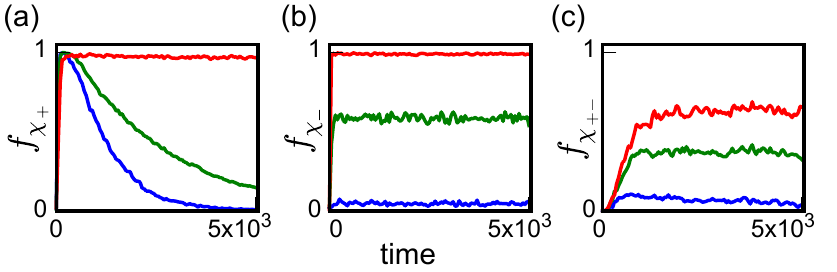}
\caption{Transient and persistent chimera states.
Fraction of (a) in-phase chimera (b) anti-chimera and (c) dual-chimera states vs. time. Mobility to coupling ratio and delay values are
(a) $\lambda /\kappa=1$ and $\tau = 1.081$ (blue line), $\tau = 1.131$ (green line) and $\tau = 1.244$ (red line); 
(b) $\lambda /\kappa=0.1$ and $\tau = 1.6$ (blue line), $\tau = 1.88$ (green line) and $\tau = 2.48$ (red line); and
(c) $\lambda /\kappa=0.001$ and $\tau = 1.36$ (blue line), $\tau = 1.44$ (green line) and $\tau = 1.6$ (red line).
The fraction is obtained over $512$ realizations for random initial conditions.
Other parameters: $\omega =1$, $N=100$. 
Parameters of the classification method: $\delta = 0.15\pi$ and $m_0=6$.}
\label{fig:xi_frac_vs_t}
\end{figure}

Using this measure we first study how the likelihood of observing chimera states changes with time, Fig.~\ref{fig:xi_frac_vs_t}.
We define the fraction $f_\chi$ as the number of realizations in which chimera states are detected by our measure, 
divided by the total number of realizations.
Starting from random initial conditions the fraction $f_{\chi_+}$ of in-phase chimera states increases rapidly and 
for some coupling delay values this fraction decays for longer times, Fig.~\ref{fig:xi_frac_vs_t}(a). 
However, there are other delay values for which in-phase chimera states persist within the time window reported here, red line in Fig.~\ref{fig:xi_frac_vs_t}(a).
The fraction $f_{\chi_-}$ of anti-chimera states quickly increases for some coupling delay values and persists with $\lambda /\kappa = 10^{-1}$, Fig.~\ref{fig:xi_frac_vs_t}(b). 
The fraction $f_{\chi_{+-}}$ of dual-chimera states grows slower, yet stays larger than zero 
for some coupling delay values with $\lambda/\kappa = 10^{-3}$, indicating their persistence, Fig.~\ref{fig:xi_frac_vs_t}(c).

\begin{figure*}[t]
\includegraphics[width=\textwidth]{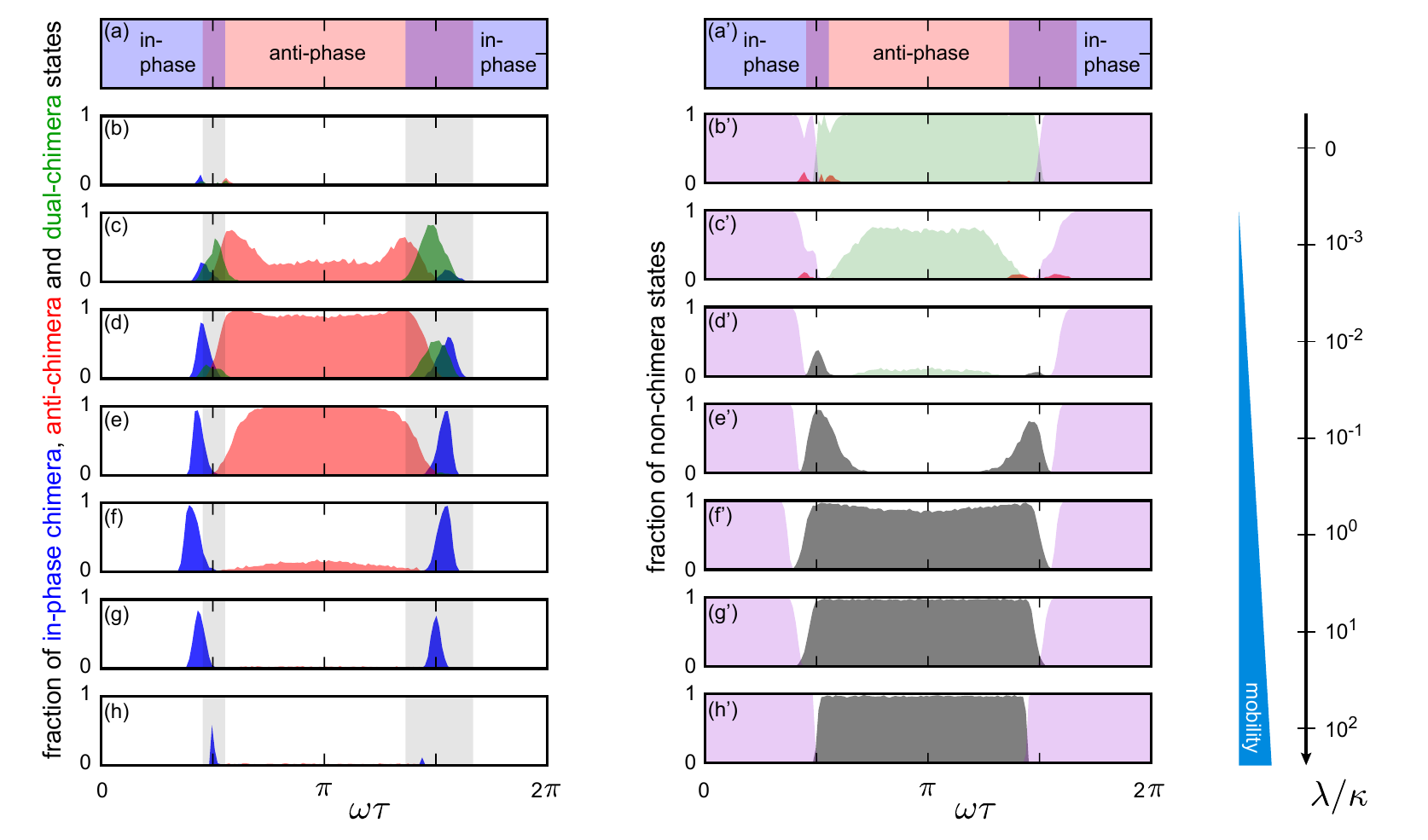}
\caption{Mobility induces chimera states from neighboring in-phase and anti-phase domains.
(a),(a') Regions in which in-phase and anti-phase order states are stable for non-mobile oscillators.
(b)-(h) Fraction of in-phase chimera (blue), anti-chimera (red) and dual-chimera states (green) as a function of coupling delay $\tau$.
Bi-stability regions of in-phase and anti-phase order states in panel (a) are replicated in the next panels as light grey bands.
(b')-(h') Fraction of in-phase local order (lilac), anti-phase local order (light green), coexistence of in-phase and anti-phase (dark red) and disordered states (dark grey) as a function of coupling delay $\tau$.
Fraction is computed at time $t=5000$, over 512 realizations of initial conditions for each set of parameters. 
Initial conditions were prepared with half the lattice with in-phase order and the other half with anti-phase order, with added Gaussian noise.
See examples of all dynamical states in Fig.~\ref{fig:signatures}.
}
\label{fig:map}
\end{figure*}
Persistent in-phase chimera states occur near the regions where in-phase and anti-phase synchronization overlap and exchange stability. 
A similar behavior has been recently observed experimentally in mechanical oscillator systems~\cite{martens13}.
%
%
For non-mobile oscillators the in-phase and anti-phase solutions of Eq.~(\ref{eq:model}) are stable 
within the regions defined by $\kappa \cos(\Omega \tau) > 0$ and $\kappa \cos(\tilde{\Omega } \tau) < 0$ respectively~\cite{nakamura94, earl03, wetzel12}, Fig.~\ref{fig:map}(a). 
In the bi-stability regions, non-mobile oscillators visit transient states where domains of in-phase and anti-phase order coexist, violet in Fig.~\ref{fig:map}(a).
Persistent chimera states occurring in the vicinity of these regions of parameter space motivated us 
to start from initial conditions that consist of separate domains of in-phase and anti-phase order.
We prepare initial conditions where half of the lattice is in-phase and the other half in anti-phase, with added Gaussian noise $\sigma = 0.1 \pi / 4$.
Such initial states do not include any disordered domains, and we can examine
whether mobility induces persistent disordered domains from ordered initial conditions, 
which would lead the system into chimera states.
We determine the fraction of chimera states that persist after a transient of $t=5000$, starting from these initial conditions.
We find persistent chimera states within an intermediate range of mobility, Fig.~\ref{fig:map}.
%
%
Without mobility persistent chimeras are not observed, Fig.~\ref{fig:map}(b). Instead, the system exhibits in-phase and anti-phase local order, Fig.~\ref{fig:map}(b').
With increasing mobility, the three different kinds of chimera states occur.
The fraction of in-phase chimera states peaks near the bi-stability regions and within the regions where in-phase order is stable.
These peaks become largest around $\lambda/\kappa=1$ and disappear for large mobility, Fig.~\ref{fig:map}(h).
Large mobility does not allow the formation of in-phase chimera states, but it rather promotes in-phase local order or disordered states depending on coupling delay values, Fig.~\ref{fig:map}(h').
Anti-chimera states form for small $\lambda/\kappa$ in the region where anti-phase order is stable,
and disappear for $\lambda/\kappa \sim 1$, Fig.~\ref{fig:map}(c-e).
Larger mobility disorganizes anti-phase structures and leads the system into disordered states, Fig.~\ref{fig:map}(f'-h').
Dual-chimera states exist in the very small mobility regime, for coupling delay values around $\pi/2$ and $3\pi/2$, Fig.~\ref{fig:map}(c,d).
All kinds of chimera states form below $\lambda/\kappa=1$, which marks the onset of the effects of mobility in systems without coupling delay, and disappear before the onset of mean field behavior~\cite{uriu13}.
We conclude that mobility induces persistent chimera states when starting from conditions that include only ordered domains.

\section{Discussion}
We have shown that in the presence of coupling delay, mobility can induce the formation of 
persistent chimera states that blend disorder with in-phase and anti-phase order in different combinations, Fig.~\ref{fig:map}.
%
%
%
%

The conditions required for chimera states to form are still matter of debate~\cite{panaggio15}.
Chimera states were first observed in systems with some form of non-local coupling~\cite{kuramoto02, abrams04, sethia08, omelchenko10},
and for some time this was thought to be a condition for chimera states to occur.
Later, chimera states were also spotted in systems with global coupling~\cite{azamat14, sethia14, schmidt14, schmidt15}. 
Chimera states in systems with local coupling have only been reported recently~\cite{laing15, bera16, clerc16, li16}. 
Here coupling is local, yet oscillators can exchange neighbors and interact with others originally far away~\cite{uriu13}.


Phase diagrams for systems of coupled mechanical oscillators reveal that chimera states appear 
in between regions where in-phase and anti-phase synchronization exchange stability~\cite{martens13}.  
Besides, chimera states are thought to occur near the stability boundaries of order and disordered states 
in systems with delayed coupling~\cite{omelchenko13, sheeba10}. 
Here we observe in-phase chimera states near the bi-stability region of in-phase and anti-phase states. 
For the very low mobility regime, in-phase chimera states form between regions where in-phase or anti-phase states dominate, Fig.~\ref{fig:map}(b')-(c'). 
For larger mobility, in-phase chimera states form between regions where in-phase order or disorder dominate, Fig.~\ref{fig:map}(d')-(h'). 
Thus, our results are consistent with both scenarios described above.

%
Analytical results for the stability of chimera states are scarce~\cite{omelchenko13, abrams08, martens16}.
Here we find that chimera states appear either as transient or persistent states, Fig.~\ref{fig:xi_frac_vs_t}.
Even the most persistent chimera states we observe are dynamic, 
with ordered domains that form and disappear in a background of disorder. 
Because mobility affects distinctly the different forms of order and disorder, there may be more than one mechanism at play.
For example mobility introduces disorder into anti-phase domains, while it favors order within in-phase domains.
The interplay of these mechanisms could underlie the different kinds of chimera states reported here.
%

%
Reliable detection and classification of chimera states poses a challenge.
%
%
Because the dynamical properties of chimera states differ between systems, different quantitative measures have been proposed to characterize chimeras~\cite{shanahan10, wolfrum11, ashwin15}.
The need for a universal systematic way of defining chimera states has motivated the development of different methods~\cite{kemeth16, gopal14, gopal15}.
%
%
%
These methods have proved to be successful in a variety of contexts, 
yet here we found the necessity to develop a new approach to distinguish between chimera states and a diversity of other dynamical states.
The computational method devised here succeeds to distinguish chimera states from every other dynamical state occurring in the system and 
furthermore allows to identify different kinds of chimera states which, to the best of our knowledge, have not been reported so far.

%
Chimeras have been recently reported in carefully designed experiments 
with photo- and electro- chemical~\cite{tinsley12, nkomo13, nkomo16, schmidt14, schonleber14}, 
optical~\cite{hagerstrom12, hart16} and mechanical systems~\cite{martens13, wojewoda16}. 
These dynamical states are thought to play a role also in some natural phenomena like unihemispheric sleep~\cite{rattenborg00}
and in psychophysical experiments~\cite{tognoli14}.
Our work suggests an unexpected avenue of research to look for chimera states in engineered and natural systems.
Technological applications featuring mobile coupled oscillators have motivated recent theoretical studies~\cite{fujiwara11, levis17}
and might give rise to chimera states provided that communication delays are present~\cite{perezdiaz17}. 
A biological system where our results could be relevant is the vertebrate segmentation clock.
%
%
The segmentation clock is a tissue generating dynamic patterns that acts during embryonic development and 
is responsible for the segmented repetitive structure of the vertebrate body axis~\cite{oates12, kageyama12, saga12, hubaud14}.
It consists of a population of genetic oscillators at the cellular level~\cite{webb16} which are coupled through a local communication mechanism~\cite{jiang00, riedel07, delaune12}.
Individual cells have to process the signals received from neighbors, cleaving and transporting macromolecules from their outer membrane to the nucleus where signals are delivered to the oscillator~\cite{wahi16}.
This introduces significant coupling delays which are thought to affect the collective rhythm and pattern formation~\cite{morelli09, herrgen10, ares12}.
Besides this delayed local coupling, cells move within the posterior part of the tissue and exchange neighbors over time~\cite{delfini05, mara07, benazeraf10, lawton13, uriu17}.
%
%
This exchange of neighbors is expected to affect information flow in the tissue~\cite{uriu10a, uriu12, uriu16, uriu17}. 
Therefore, both key ingredients in the theory are present in this system and it is possible that perturbations to delays or mobility could induce the formation of chimera states.

\vspace{0.2cm} \noindent
\emph{Acknowledgments.} 
We thank I. M. Lengyel and J. N. Freitas for valuable comments on the manuscript.
The authors would like to thank Centro de Simulaci\'on Computacional para Aplicaciones Tecnol\'ogicas (CSC-CONICET) for computational resources.
LGM acknowledges support from ANPCyT PICT 2012 1954 and PICT 2013 1301, and FOCEM-Mercosur (COF 03/11).
KU acknowledges support from JSPS KAKENHI Grant Number 26840085, 
FY2014 Researcher Exchange Program between JSPS and CONICET and Kanazawa University Discovery Initiative program.

\section*{Appendix A: Numerical methods}
As described in the main text 
we consider a system of $N$ phase oscillators with phases $\theta_i $, with $i=1, \ldots , N$, placed at discrete positions $x_i=1,\ldots, N$ in a one dimensional lattice of $N$ sites. 
The phases of the oscillators evolve according to Eq.~(\ref{eq:model}), which we integrate numerically using a fourth order Runge-Kutta algorithm with a constant time step $dt$.
Eq.~(\ref{eq:model}) includes a delayed coupling between first neighbors in the lattice.
The value of the delay in the coupling is one of the relevant parameters of the model and is always set as an integer multiple of the time step
\begin{equation} \label{eq:tau}
\tau = n_\tau \, dt
\end{equation}
where $n _\tau $ is a natural number.

Additionally, oscillators are able to move through the lattice by exchanging positions with their nearest-neighbors.
%
%
%
This stochastic exchange of oscillators positions is described as a Poisson process.
We introduce a mobility rate $\lambda$ so that each pair of neighboring oscillators has a probability $\lambda /2$ of exchanging positions per unit time.
The mobility rate $\lambda $ is another relevant parameter of our model and can be set independently from the other parameters, such as the delay value $\tau $.

To simulate this Poisson process we use an approximation of the Gillespie algorithm for discrete time intervals.
%
The distribution of waiting times $t_e$ for the next exchange event in the lattice of $N$ oscillators is
\begin{equation} \label{eq:flipping_times}
P(t_e)= a_0 \, \exp{(-a_0 \, t_e)}
\end{equation}
where $a_0 = (N-1) \, \lambda/2$ is the propensity for an exchange event~\cite{gillespie76}. 
We generate a discrete set of random waiting times $\{ t _e \}$ drawn from this distribution. 
%
%
We approximate each of the waiting times $t_e$ in this set by the closest larger integer multiple of $dt$. 
As a result we obtain a discretized waiting time $t_{ed}$
\begin{equation}
t_e \rightarrow t_{ed} = n_e \, dt \: \: \: \: \text{with} \: \: \: \: n_e = \text{ceil}(t_e/dt).
\end{equation}
%
%
%
Therefore, exchange events occur at time points that are integer multiples of the integration step.
This discretization of the waiting times introduces a perturbation from the Poissonian statistics. 
For this perturbation to be negligible, this approximation requires small enough $dt$ compared to the average time interval between two successive exchange events $\langle t_e \rangle =1/a_0$, that is $dt << 1/a_0$.
Thus, to obtain accurate realizations of exchange events we set $dt \leqslant {\langle t_e \rangle}/{10}$ in all simulations~\cite{uriu12}.
If $dt > 0.01 $ in this equation, we set $dt=0.01$.
Thus, the time step $dt$ for numerical integration is fixed within each simulation and is the same for simulations with the same parameter values. 

In summary, the theoretical model described here is implemented numerically as follows. 
Given a mobility rate $\lambda $ we fix a time step $dt$ as stated before.
Next we set the delay value $\tau$ as in Eq.~(\ref{eq:tau}) by choosing a number $n_\tau $ of time steps.
We then choose the initial phases $\theta _i(t=0)$ and positions $x_i(t=0) $ for the oscillators $i=1, \ldots N$ and 
set their history for $t$ in $[-\tau, 0)$. 
No exchange events occur before $t=0$, so $x_i(t)=x_i(t=0)$ for $t$ in $[-\tau, 0)$ and $\forall i$. 
Finally, we iterate the following steps:
\begin{enumerate}
\item Generate the time for the next exchange event $t_e$ from the distribution in Eq.~(\ref{eq:flipping_times}) and also randomly choose which pair of oscillators will exchange positions from a uniform distribution in the amount of nearest-neighbor pairs $\mathcal{U}(N-1)$.
\item Integrate $n_e$ time steps of Eq.~(\ref{eq:model}) with Runge-Kutta algorithm, with fixed positions for all the oscillators in the lattice.
\item Update positions and neighborhoods affected by the exchange event and go back to step 1.
\end{enumerate}

\section*{Appendix B: Classification of dynamical states}
To measure a fraction of chimera states and study how this fraction changes with time and with parameter values, 
we need to distinguish between the different kinds of states that occur.
There is not a unique method for discriminating chimeras 
which is useful for the vast variety of systems where they occur~\cite{gopal14, gopal15, kemeth16}. 
\begin{figure*}[b]
  \centering
\includegraphics[width=\textwidth]{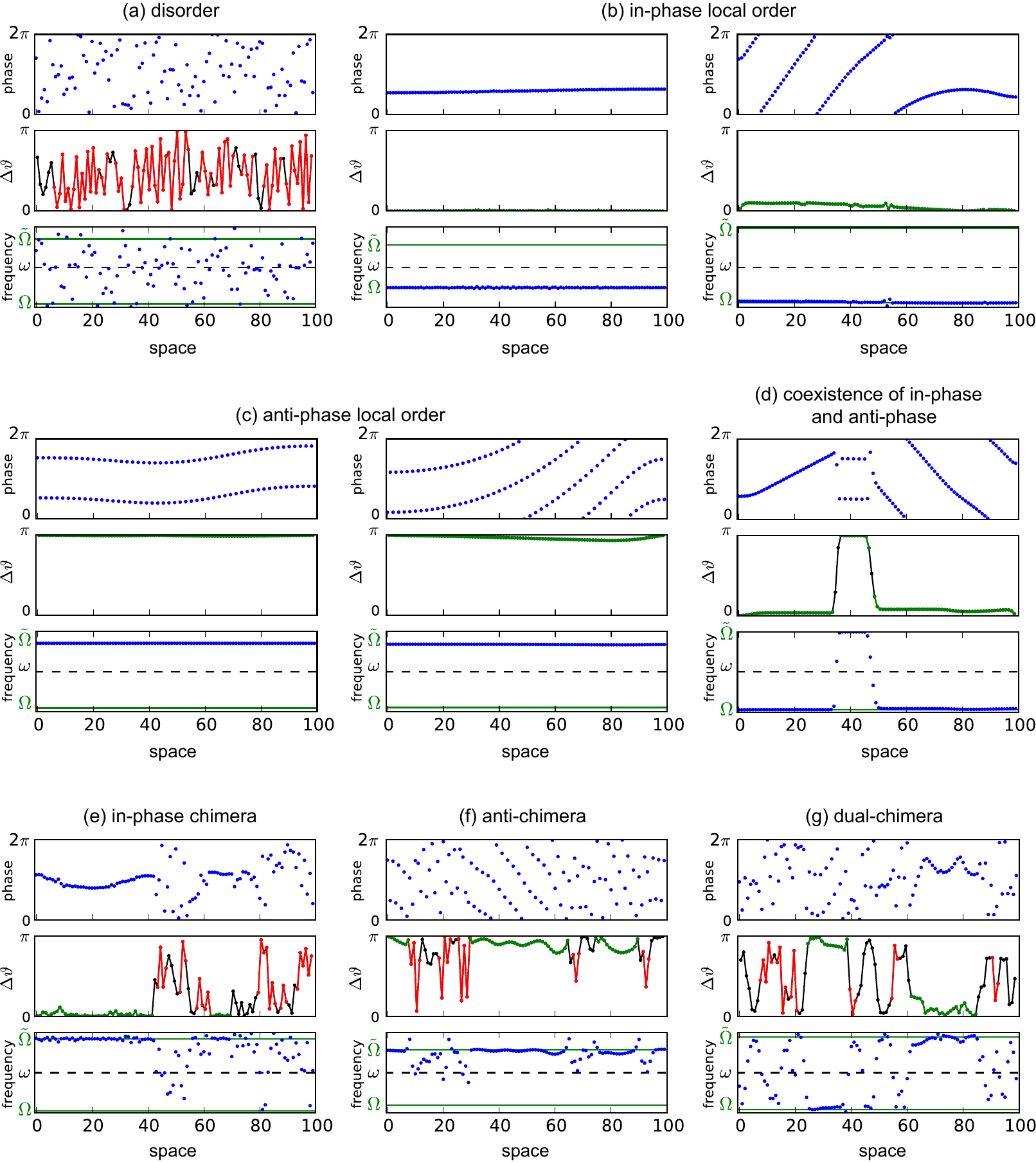}
\caption{\label{fig:signatures}
Examples of the variety of dynamical states that the system can visit depending on parameters. For each of the states, top, middle and bottom panels show a snapshot of phases, the spatial profile for phase differences, and instantaneous phase velocities, respectively.
Middle panel illustrates how the classification method works for $\delta=0.15\pi$ and $m_0=6 \,$: peaked motifs are marked in red and ordered domains are green. 
Solid green lines in the bottom panel show the in-phase $\Omega$  and anti-phase $\tilde{\Omega}$ collective frequencies. Parameters: (a) $\lambda / \kappa=10^2$ and $\tau=2.2$, (b) $\lambda / \kappa=10$ and $\tau=0.56$ (left), $\lambda / \kappa=0.01$ and $\tau=1.2$ (right), (c) $\lambda / \kappa=0$ and $\tau=2.2$ (left), $\lambda / \kappa=0$ and $\tau= 2.24$ (right), (d) $\lambda / \kappa=0$ and $\tau=1.36$, (e) $\lambda / \kappa=1$ and $\tau=4.88$, (f) $\lambda / \kappa=0.01$ and $\tau=2.4$, (g) $\lambda / \kappa=0.01$ and $\tau=4.76$.}
\end{figure*}
In the system considered here, a diversity of dynamical states are present together with chimera states, Fig.~\ref{fig:signatures}(a-g) top panels:
\begin{enumerate}[label=(\alph*)]
\item disorder
\item in-phase local order
\item anti-phase local order
\item coexistence of in-phase and anti-phase
\item in-phase chimera states ($\chi_+$)
\item anti-chimera states ($\chi _-$)
\item dual-chimera states ($\chi _{+-}$)
\end{enumerate}

\begin{figure}[t]
\includegraphics[width=\columnwidth]{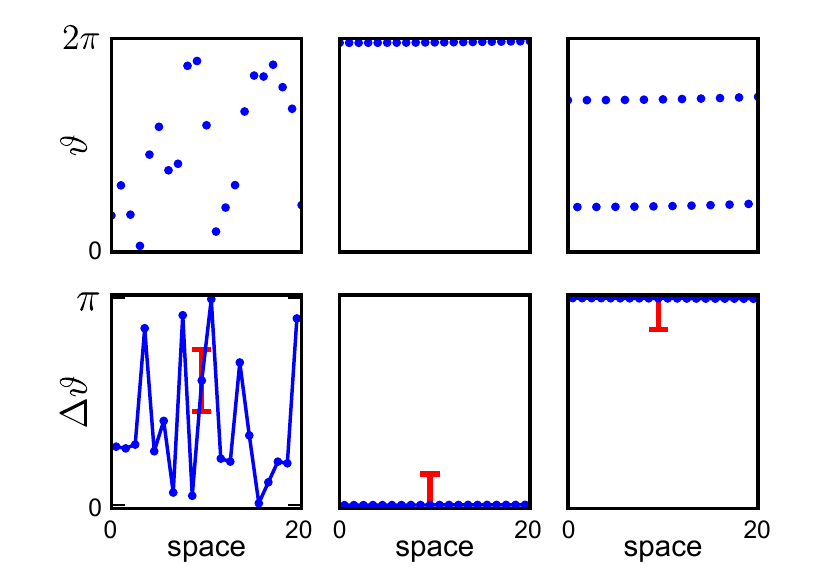}{\caption{\label{fig:ord_dis_motifs} 
Typical disorder and order phase snapshots (top panel) and corresponding phase differences (bottom panel). 
Red error bars indicate the noise tolerance window defined by $\pm \delta$, here $\delta=0.15\pi$.
For ordered patterns all subsequent phase differences lay inside the noise tolerance window 
while most subsequent phase differences in the disordered pattern lay outside.}
}
\end{figure}

Our approach uses phase differences to identify ordered and disordered domains in the lattice.
Given a snapshot at time $t$ of the system state, we first consider the absolute value of phase differences between first neighbours, modulo $2\pi$:
$$
\Delta \vartheta _k(t) = \min \left\lbrace \, \vert \vartheta _{k+1}(t) - \vartheta _k(t) \vert \, , \, 2\pi - \vert \vartheta _{k+1}(t) - \vartheta _k(t) \vert \, \right\rbrace 
$$
where $\vartheta _k(t)$ is the phase value at site $k$ at the time $t$, with $k=0,...,N-2$, Fig.~\ref{fig:ord_dis_motifs} bottom panels.
While disordered parts of the snapshots display variable phase differences with large changes from one site to the next, phase differences for ordered domains remain almost constant, Fig.~\ref{fig:ord_dis_motifs}.
Therefore, we seek a way to identify whether consecutive phase differences change abruptly going up or down, or stay almost constant.

\begin{figure}[b]
\includegraphics[width=1\columnwidth ]{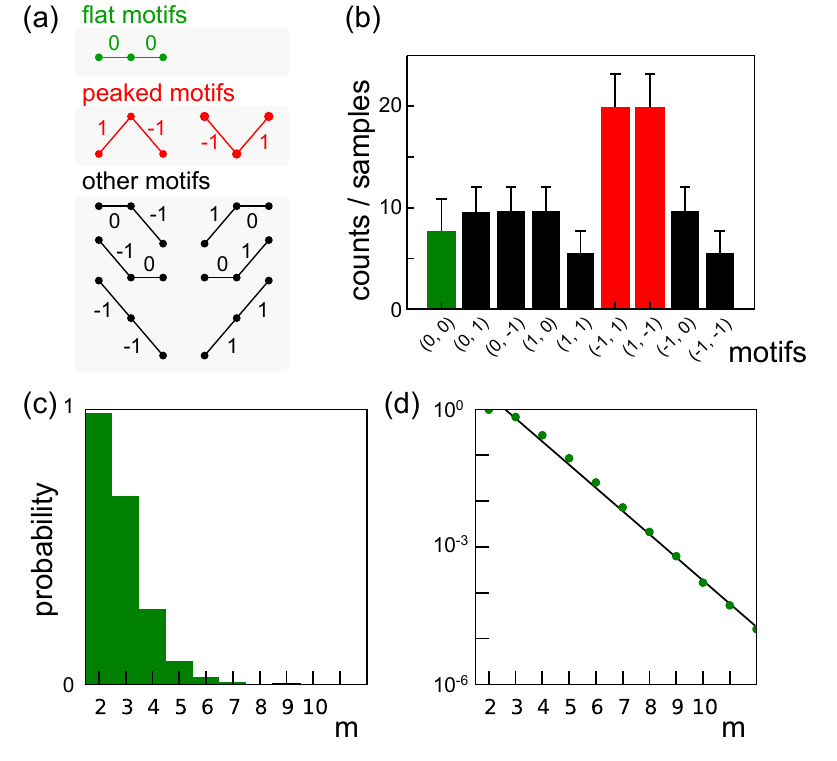}
\caption{
(a) Possible phase difference motifs arranged to show the signatures of order (top) and disorder (middle). 
Other occurring motifs (bottom) are not used by our classification scheme.
(b) Histogram of three-node motifs for disordered states. Green and red bars are flat and peaked motifs, respectively, black bars are all other motifs. Snapshots were prepared by taking $N=100$ phases from a uniform distribution between $[0, 2\pi]$. Total sample snapshots considered: $5\times 10^5$.
(c, d) Probability of finding at least one domain of $m$ consecutive zeros in a disordered state. 
Probability was computed over $10^6$ sample disordered snapshots, with noise threshold $\delta =0.15\pi$. 
Black line in right panel shows the exponential fit used to estimate the decay rate and to choose $m_0$.}
\label{fig:motifs}
\end{figure}

We introduce phase difference motifs consisting of three nodes, corresponding to three consecutive phase differences, Fig.~\ref{fig:motifs}(a).
Motifs are labeled with two numbers, one for each of its two links. 
These numbers reflect how similar a phase difference $\Delta \vartheta _k$ is from the following $\Delta \vartheta _{k+1}$.
We assign the labels to a link according to the following criteria:
\begin{itemize}
\item if $-\delta < \Delta \vartheta _{k+1} - \Delta \vartheta _k < \delta $, link value is $0$
\item if $\Delta \vartheta _{k+1} - \Delta \vartheta _k > \delta $, link value is $1$
\item if $\Delta \vartheta _{k+1} - \Delta \vartheta _k < - \delta $, link value is $-1$ \,.
\end{itemize}
The quantity $\delta$ determines the threshold of noise that we admit for defining order and is a parameter of our method.
%
%
In Figs.~\ref{fig:xi_frac_vs_t} and \ref{fig:map} we choose $\delta =0.15\pi$, which is a $15\%$ of the maximum possible value of the phase differences.

Fig.~\ref{fig:motifs}(b) shows the distribution of motifs for a snapshot consisting of randomly chosen phases for all the oscillators in the one dimensional lattice. 
It becomes evident that such disordered snapshots are characterized by a larger fraction of peaked motifs $\lbrace (1,-1), (-1,1)\rbrace $ than other motifs.
Thus, peaked motifs are a hallmark of disorder and we consider the presence of at least one peaked motif in a snapshot as an indicator that some amount of disorder is present in the system.

Similarly, we can identify the presence of order by looking for flat motifs $\lbrace (0,0) \rbrace $. 
Flat motifs can also happen by chance in disordered states, Fig.~\ref{fig:motifs}(b). 
Therefore we consider that there is order present in the system if there is at least one domain with a minimum amount $m_0$ of consecutive zeros. 
To calibrate this parameter, we study the distribution of consecutive zeros in disordered states, Fig.~\ref{fig:motifs}(c,d).
The number of consecutive zeros that could appear in a disordered state decays exponentially. 
We consider as a reference the value of $m$ for which the exponential falls to a value of $1\% $ of its maximum for $m=2$. 
A linear fit shows this happens roughly for $m>5$, Fig.~\ref{fig:motifs}(d). 
Then, we consider that if at least $m_0=6$ consecutive zeros are present in a snapshot of the system state, the snapshot presents an ordered domain. 
This domain could have the size of the system or could coexist with other motifs.

With the described procedure, we are able to locally distinguish the presence of order and disorder in a snapshot of the system state, Fig.~\ref{fig:signatures}(a-g) middle panels. 
When domains with at least $m_0$ consecutive zeros coexist with at least one peaked motif, our approach identifies a chimera state. 
In-phase and anti-phase ordered domains can be distinguished by evaluating the mean value of the phase differences within the domains.
Therefore, our approach is capable to classify the seven types of states displayed in Fig.~\ref{fig:signatures}, 
according to which domain kinds are present.


\end{document}